\documentclass[twocolumn,british,pra,aps,showpacs,superscriptaddress]{revtex4-2}
\usepackage[letterpaper]{geometry}
\usepackage{url}
\usepackage{graphicx}
\usepackage{amsmath,bbm}
\usepackage{amssymb}
\usepackage{esint}
\usepackage{epstopdf}
\usepackage{color}
\usepackage{babel}
\usepackage{graphics}
\usepackage{epsfig}
\usepackage{bbold}
\newcommand{\Tr}{\textrm{Tr}}
\newcommand{\ket}[2][]{{|#2\rangle_{#1}}}
\newcommand{\bra}[2][]{{}_{#1}\langle #2|}
\newcommand{\braket}[3][]{{{}_{#1}\langle#2|#3\rangle_{#1}}}


\newcommand{\eqnref}[1]{eq.~(\ref{#1})}
\newcommand{\eqnsref}[2]{eqs.~(\ref{#1})-(\ref{#2})}

\begin{document}

\title{Quantum limits to polarization measurement of classical light}
\author{Marcin Jarzyna}
\affiliation{Centre for Quantum Optical Technologies, Centre of New Technologies, University of Warsaw, ul. Banacha 2c, 02-097 Warszawa}
\date{\today}

\begin{abstract}
Polarization of light is one of the fundamental concepts in optics. There are many ways to measure and characterise this feature of light but at the fundamental level it is quantum mechanics that imposes ultimate limits to such measurements. Here, I calculate the quantum limit to a precision of a polarization measurement of classical coherent light. This is a multiparameter estimation problem with a crucial feature of noncommuting optimal observables corresponding to each parameter which prohibits them to be measured at the same time. I explicitly minimize the quantum  Holevo-Cramer-Rao bound which tackles this issue and show that it can be locally saturated by two types of conventional receivers.
\end{abstract}

\maketitle

\section{Introduction}

Polarization measurement lies at the core of many scientific and technical endeavors. It is a standard venture in many applications in communication \cite{Han2005, Martinelli2006, Zhou2009, Rosskopf2020, Kikuchi2020}, metrology \cite{Angelsky2009, Salvail2013, Toeppel2014, Goldberg2021} and imaging \cite{Demos1997, Colomb2002, Sattar2020}. Increasing precision of light polarization measurements is of crucial importance for enhancing performance in various aspects of these fields. In order to accomplish this goal one may employ different detection schemes and eliminate various kinds of systematic errors and noise. However, at the fundamental level both light and the detection process are described by the principles of quantum mechanics which is known to impose ultimate limits on the estimation precision for a variety of physical parameters like optical phase, magnetic field strength, light source separation etc. \cite{Giovannetti2006, DemkowiczDobrzanski2012, Baumgratz2016, Tsang2016}. It should be therefore expected that similar bounds can be constructed for the polarization estimation.

The issue of finding ultimate limits to polarization measurement precision has been extensively studied in the literature. For single photon states, which polarization states can be described by a qubit, itself characterized by the position of a point in the Bloch ball, the bounds were found in parameter estimation and tomographic pictures \cite{Rehacek2004, Bagan2006, Ling2006, Salvail2013}. More general states of light and approaches have been considered recently \cite{Klimov2010, Rudnicki2020, Norrman2020, Martin2020, Goldberg2021, Goldberg2021a} indicating plethora of interesting results, such as precision enhancement in the presence of entanglement \cite{Toeppel2014}. Still, however, the exact ultimate bounds on the precision of polarization measurement for many important instances have not been properly studied or are unknown at all.

Polarization estimation is in general an example of a multiparameter estimation problem. As such, a proper care needs to be taken in order to find meaningful bounds on estimation precision. In particular, the multiparameter quantum Cramer-Rao bound (QCRB) \cite{Helstrom1976, Braunstein1994} often used in the literature to ascertain ultimate precision limits allowed by quantum mechanics is known to be in fact not attainable in general. This is because in quantum picture, unlike its classical counterpart, measurements of different physical quantities may not commute with each other, preventing them to be performed simultaneously on the same system. Another issue is the assumption of resources that may not be present in practical realizations, such as global phase reference. Lastly, the bounds itself may just serve as benchmarks of optimal performance but it is an equally vital problem to identify realistic measurement schemes that allow to saturate them.  It is therefore of crucial importance to use proper tools, like Holevo-Cramer-Rao bound (HCRB) \cite{Holevo1982}, and keep in mind physical limitations of realistic systems in order to find actual saturable bounds on precision and means to attain them.

In this paper I consider polarization measurements of a classical light. The latter is conventionally characterized by the amplitudes of electromagnetic field in two orthogonal polarization modes. In the quantum picture a classical light state with arbitrary polarization is described by a two mode coherent state $\ket{\alpha_H}_H\otimes\ket{\alpha_V}_V$, where $\alpha_H,\,\alpha_V$ denote complex amplitudes in horizontal and vertical polarization modes $H$, $V$ respectively. The exact state of polarization depends on the absolute values of the amplitudes and the relative phase between them. This description covers all fully polarized classical states. One may also consider partially polarized light which includes an incoherent admixture, however, this is outside the scope of this work.

\begin{figure}
	\centering
	\includegraphics[width=0.8\columnwidth]{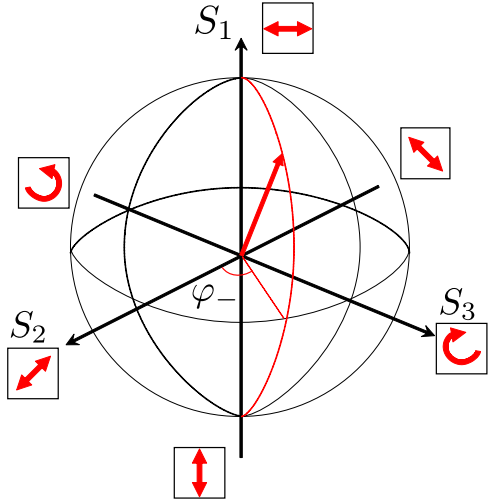}
	\caption{Poincare ball representing states of polarization. For fully polarized light $S_0$ is the radius of the ball. Poles on axes $S_1,\,S_2,\,S_3$ represent respectively states of vertical-horizontal, diagonal-antidiagonal and right -left circular polarized light, as indicated by the arrows. $\varphi_-=\varphi_H-\varphi_V$ denotes relative phase between the polarization modes.}
\label{fig:poincare}
\end{figure}

Polarization of light is usually described in terms of the Stokes vector $\vec{S}=(S_0,S_1,S_2,S_3)$ \cite{Stokes1852, Born1985} which is illustrated in Fig.~\ref{fig:poincare}. For the fully polarized classical light the coordinates $S_j$ can be expressed by the amplitudes $\alpha_{H,V}$ as
\begin{gather}
S_0=|\alpha_H|^2+|\alpha_V|^2,\nonumber\\
S_1=|\alpha_H|^2-|\alpha_V|^2,\nonumber\\
S_2=\alpha_H^*\alpha_V+\alpha_V^*\alpha_H,\nonumber\\
S_3=i(\alpha_H^*\alpha_V-\alpha_V^*\alpha_H).\label{eq:stokes_parameter3}
\end{gather}
Note that $S_0=\sqrt{S_1^2+S_2^2+S_3^2}$ and it is proportional to the average energy carried by the light field. Horizontal and vertical polarization states are obtained for $S_1>0$ and $S_1<0$ respectively while $S_2,\,S_3=0$. Similarly diagonal and antidiagonal polarization occurs whenever $S_2>0$ or $S_2<0$ with $S_1,\,S_3=0$ and the light is left or right circularly polarized for $S_3>0$ and $S_3<0$ with $S_1,\,S_2=0$ respectively. Various other kinds of elliptical polarization states may be obtained in intermediate scenarios.

In the quantum picture Stokes vector coordinates are promoted to Stokes operators \cite{Jauch1980, Korolkova2002} given by
\begin{gather}
\hat{S}_0=\hat{a}^\dagger\hat{a}+\hat{b}^\dagger\hat{b},\nonumber\\
\hat{S}_1=\hat{a}^\dagger\hat{a}-\hat{b}^\dagger\hat{b},\nonumber\\
\hat{S}_2=\hat{a}^\dagger\hat{b}+\hat{b}^\dagger\hat{a},\nonumber\\
\hat{S}_3=i(\hat{a}^\dagger\hat{b}-\hat{b}^\dagger\hat{a}),\label{eq:stokes_operator3}
\end{gather}
where $\hat{a},\hat{a}^\dagger$ and $\hat{b},\hat{b}^\dagger$ denote annihilation and creation operators for the horizontal and vertical light modes respectively. The Stokes parameters in eqs.~(\ref{eq:stokes_parameter3}) are obtained as expectation values of respective observables in eqs.~(\ref{eq:stokes_operator3}) on the quantum state of light $\ket{\alpha_H}\ket{\alpha_V}$. Each such observable has a non-vanishing variance $\Delta S_j^2=S_0$, meaning that polarization components cannot be measured exactly. Crucially, this is not the only limitation set by quantum mechanics on the polarization measurement. This is because observables $\hat{S}_j$ for $j=1,2,3$ do not commute with each other, $[\hat{S}_j,\hat{S}_k]=2i\epsilon_{jkl} \hat{S}_l$, which means that they cannot be measured simultaneously on a given quantum state. Therefore, in general a naive bound on the precision $\Delta S_1^2+\Delta S_2^2+\Delta S_3^2\geq 3S_0$ which one could write using just uncertainties of each component cannot be saturated. Finding the actual attainable lower bound on the precision is the main goal of this manuscript.

\section{Quantum estimation theory}

As mentioned in the previous section, because of noncommutativity of the Stokes operators one cannot simply measure polarization directly and hope for optimal precision. Instead, a detection of some particular quantity, described by a positive operator valued measure (POVM) $\{\Pi_y\}$, needs to be performed first. Next, from the results of the measurement an estimate of the Stokes parameters can be found through estimators $\tilde{S}_j(y),\,j=1,2,3$. In order to find the best estimation precision allowed by the laws of quantum mechanics it is then necessary to optimize the precision over the POVMs and estimators. This task seems difficult at the first glance, but it can be vastly simplified by using tools from quantum estimation theory.

In a general quantum estimation problem \cite{Helstrom1976} one is concerned about estimating a set of parameters $\theta_j$ combined in a vector $\pmb{\theta}=(\theta_1,\theta_2,\dots)$ from a parameter-dependent quantum state $\rho_{\pmb{\theta}}$. For a particular measurement described by a POVM $\Pi_y$ performed on the state the distribution of outcomes is given by $p(y|\pmb{\theta})=\Tr(\rho_{\pmb{\theta}}\Pi_y)$. In order to obtain estimates of the parameter values one utilizes an estimator function $\tilde{\pmb{\theta}}(y)=(\tilde{\theta}_1(y),\tilde{\theta}_2(y),\dots)$. The precision of such procedure is quantified by the cost $\Delta\tilde{\pmb{\theta}}^2=\sum_j\Delta\tilde{\theta}_j^2$ i.e. sum of mean squared errors for each parameter $\Delta\tilde{\theta}_j^2=\sum_y p(y|\pmb{\theta})(\tilde{\theta}_j(y)-\theta_j)^2$. In such setting the lower bound on precision for any unbiased estimator is given by the Cramer-Rao bound \cite{Kay1993}
\begin{equation}\label{eq:class_CR}
\Delta\tilde{\pmb{\theta}}^2\geq \Tr \left(\mathcal{F}^{-1}\right),
\end{equation}
where $\mathcal{F}$ is called the Fisher information matrix and equal to
\begin{equation}\label{eq:class_FI}
\mathcal{F}_{jk}=\sum_y\frac{1}{p(y|\pmb{\theta})}\frac{\partial p(y|\pmb{\theta})}{\partial\theta_j}\frac{\partial p(y|\pmb{\theta})}{\partial\theta_k}.
\end{equation}
This bound can be in principle saturated by a maximum likelihood estimator.

Importantly, the bound in \eqnref{eq:class_CR}, is optimized only over estimators and depends on the measurement. Therefore, in order to find the best possible precision one needs to additionally maximize the Fisher information matrix over the POVMs. This is a hard task, so in order to avoid this cumbersome optimization one often uses a quantum Cramer-Rao bound (QCRB)\cite{Helstrom1976, Braunstein1994}
\begin{equation}\label{eq:quantum_CR}
\Delta\tilde{\pmb{\theta}}^2\geq \Tr\left(\mathcal{F}^{-1}\right)\geq \Tr\left(\mathcal{Q}^{-1}\right),
\end{equation}
where $Q$ is the quantum Fisher information matrix equal to
\begin{equation}\label{eq:quantum_FI}
\mathcal{Q}_{jk}=\frac{1}{2}\Tr[\rho_{\pmb{\theta}}(L_j L_k+L_k L_j)].
\end{equation}
Operators $L_j$ are called symmetric logarithmic derivatives (SLD) and are implicitly defined as solutions to the equation
\begin{equation}
\frac{\partial \rho_{\pmb{\theta}}}{\partial\theta_j}=\frac{1}{2}(\rho_{\pmb{\theta}}L_j+L_j\rho_{\pmb{\theta}}).
\end{equation}
The quantum Cramer-Rao bound for single parameter problems, i.e. when $\pmb{\theta}=\theta$, may always be saturated by performing a projective measurement in the eigenbasis of SLD operators. However, if there is more than one parameter, finding the optimal POVM posses a challenge since if any two SLDs do not commute with each other then it is not possible to find their common eigenbasis. Therefore the bound in \eqnref{eq:quantum_CR} is in general not attainable.

The actual saturable lower bound for multiparameter quantum estimation problem is given by the Holevo-Cramer-Rao bound (HCRB) \cite{Holevo1982, Nagaoka1989, Ragy2016, Demkowicz2020}, which unfortunately does not enjoy any kind of elegant and simple formula like the QCRB in \eqnref{eq:quantum_FI}. The expression for HCRB reads
\begin{equation}\label{eq:HCRB}
\Delta\tilde{\pmb{\theta}}^2\geq \min_{\{X_j\}}\{\Tr (\textrm{Re} Z[X])+||\textrm{Im}Z[X]||_1\},
\end{equation}
where the matrix $Z[X]_{jk}=\Tr(\rho_{\pmb{\theta}}X_j X_k)$, $||A||_1=\Tr\sqrt{A^\dagger A}$ is the trace norm and the minimization is performed over all Hermitian matrices satisfying $\Tr(X_j\partial_{\theta_k}\rho_{\pmb{\theta}})=\delta_{jk}$ and $\Tr(X_j\rho_{\pmb{\theta}})=0$. The QCRB and HCRB are equivalent if and only if the commutators of all SLDs evaluated on the state vanish $\Tr(\rho_{\pmb{\theta}}[L_j,L_k])=0$. It is known that QCRB results in a bound that is at most twice as large as HCRB \cite{Carollo2019, Tsang2020}. Importantly, if one is interested in estimating quantities $\pmb{\theta}'$ which are given by a function of the original parameters $\pmb{\theta}'=\pmb{f}(\pmb{\theta})$ the HCRB transforms according to the rules $X'=JX$ and $Z[X]'=J Z[X] J^T$ where $J_{jk}=\partial \theta_j'/\partial \theta_k$ is the Jacobian matrix.

A crucial feature of the HCRB bound is that it takes care of the issue of noncommutativity of the optimal observables corresponding to different parameters. A profound consequence of this fact is the existence of the so called nuisance parameters \cite{Yang2019, Suzuki2020, Suzuki2020a, Demkowicz2020}. These are any unobserved, i.e. not estimated, parameters in the model for which corresponding observables do not commute with the ones related to the estimated quantities. Because of the noncommutativity, the existence of nuisance parameters can affect the form of the optimal observable for the observed parameters and the precision bound for them. In multiparameter estimation problems it is therefore important to carefully analyze all the parameters describing the state, even if they are not being estimated in a particular scenario, in order to find valid bounds on precision.

\section{HCRB for polarization estimation}
\label{sec:hcrb}

In the case of polarization measurement the Stokes parameters in \eqnref{eq:stokes_parameter3} depend on the amplitudes $\alpha_H,\,\alpha_V$ of the polarization modes. One could therefore naively conclude that an equivalent estimation problem is to estimate both complex amplitudes. The optimal measurement for the latter problem is known to be a double homodyne measurement of each polarization mode \cite{Holevo1982}. The bound on the polarization estimation precision can be then shown to be equal to $6S_0$ \cite{Kikuchi2020, Mecozzi2021}, twice as much as would follow from separate optimal measurement of each Stokes parameter and summing up the precisions. However, by writing coherent states amplitudes explicitly as $\alpha_H=|\alpha_H| e^{i\varphi_H}$ and $\alpha_V=|\alpha_V|e^{i\varphi_V}$ the Stokes parameters can be expressed as
\begin{gather}
S_0=|\alpha_H|^2+|\alpha_V|^2,\nonumber\\
S_1=|\alpha_H|^2-|\alpha_V|^2,\nonumber\\
S_2=2|\alpha_H||\alpha_V|\cos\varphi_-,\nonumber\\
S_3=2|\alpha_H||\alpha_V|\sin\varphi_-,\label{eq:stokes_parameter_ph3}
\end{gather}
where $\varphi_-=\varphi_H-\varphi_V$ is the relative phase between the polarization modes. It can be seen in eqs.~(\ref{eq:stokes_parameter_ph3}), that only the absolute values of the amplitudes and the relative phase between the modes enter the expression for the Stokes parameters. One can therefore rephrase estimation of $\vec{S}$ as estimation of just three real parameters $(|\alpha_H|,|\alpha_V|,\varphi_-)$ rather than the full complex amplitudes $\alpha_H,\alpha_V$. On the other hand, in order to fully describe the quantum state $\ket{\alpha_H}_H\otimes\ket{\alpha_V}_V$ one needs to add to those three quantities also the global phase $\varphi_+=\varphi_H+\varphi_V$. It is well known that amplitude, or more likely the number of photons, and phase exhibit a Heisenberg-like uncertainty relation \cite{Smithey1993}, meaning that their corresponding observables are not commuting. This strongly suggests that $\varphi_+$ is a nontrivial nuisance parameter in the polarization estimation problem and that double homodyne detection is in principle suboptimal for polarization measurement. Another important problem is that in order to meaningfully consider global phase one needs to posses an external reference beam with respect to which this phase is defined \cite{Molmer1997, Jarzyna2012}.

The detailed calculations leading to HCRB on polarization estimation precision are quite complicated and given in the appendix \ref{app:hcrb}. Importantly, the inclusion of a global phase as an unobserved nuisance parameter manifests in an additional constraint on the form of matrices $X_j$ in \eqnref{eq:HCRB}
\begin{equation}\label{eq:XH0const1}
\textrm{Re}\left[\bra{\alpha_H}\bra{\alpha_V}X_j\,\partial_{\varphi_+}\left(\ket{\alpha_H}\ket{\alpha_V}\right)\right]=0,
\end{equation}
for $j=|\alpha_H|,|\alpha_V|,\varphi_-$. The presence of the above constraint completely fixes the form of matrices $X_j$, removing the need of optimization in\eqnref{eq:HCRB}. The final bound, after rephrasing it in terms of Stokes coefficients reads
\begin{equation}\label{eq:HCRB_pol}
\Delta S_1^2+\Delta S_2^2 +\Delta S_3^2\geq 5S_0.
\end{equation}
Crucially, the above expression lies below the value of $6S_0$ attainable by double homodyne detection and is the lower limit imposed by the laws of quantum mechanics. The limit in \eqnref{eq:HCRB_pol} has been conjectured recently in \cite{Mecozzi2021}, where a concrete estimation scheme locally saturating this bound was found. Note also, that the value of HCRB in \eqnref{eq:HCRB_pol} firmly lies in the region allowed by the QCRB $\Tr(\mathcal{Q}^{-1})\leq \Tr(C_{HCRB})\leq 2\Tr(\mathcal{Q}^{-1})$ with $\Tr(\mathcal{Q}^{-1})=3S_0$, as discussed in appendix~\ref{app:qfi}.

An important instance of polarization estimation is the case in which the total power, $S_0$, is known. In such a case one can use the fact that $|\alpha_V|=\sqrt{S_0-|\alpha_H|^2}$ which means that there are only two parameters to estimate $(|\alpha_H|,\,\varphi_-)$, still, however, one must include global phase as a nuisance parameter. The precision bound in this scenario is given by
\begin{equation}\label{eq:HCRB_const}
\Delta S_1^2+\Delta S_2^2+\Delta S_3^2\geq4S_0.
\end{equation}
The detailed derivation is presented in appendix \ref{app:hcrb_s0}. This bound is lower than the one in \eqnref{eq:HCRB_pol} which is not surprising since one estimates only two instead of three parameters. It is also two times worse than the corresponding QCRB, which in turn is equal to $\Tr(\mathcal{Q}^{-1})=2S_0$, see Appendix~\ref{app:qfi}. 

\section{Attaining the HCRB}
\label{sec:stokes}

\begin{figure}
	\centering
	\includegraphics[width=\columnwidth]{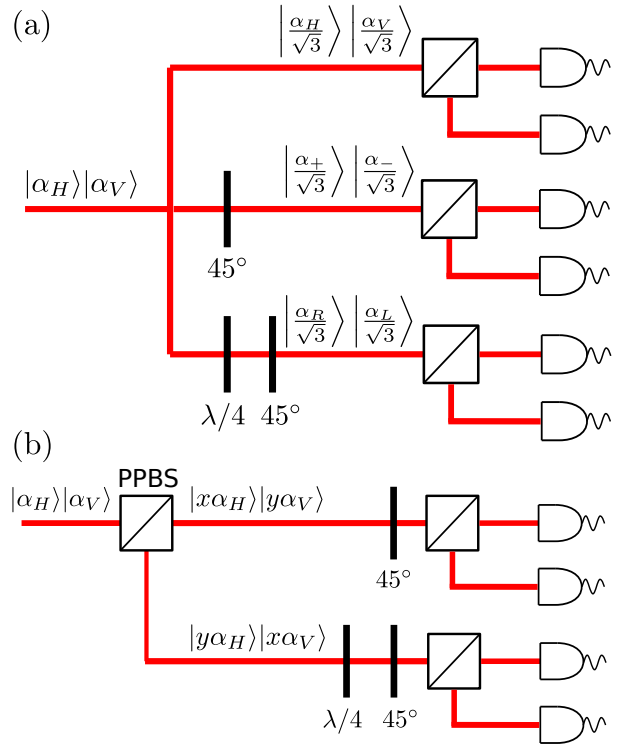}
	\caption{Schemes of Stokes (a) and tetrahedron (b) detection schemes. Here $\alpha_\pm$ and $\alpha_{R,L}$ represent amplitudes in diagonal-antidiagonal and right-left circular polarization bases.}
\label{fig:stokes}
\end{figure}

A nontrivial question in many estimation tasks is what is the measurement that saturates the HCRB bound. This problem has a solution for a general estimation procedure for pure states \cite{Matsumoto2002}, however, the resulting POVM is often of limited practical use as it is complicated and its physical realization is rather difficult to obtain. In this work I will take a different approach, that is, I will consider a specific POVM and show that it attains the HCRB bound in \eqnref{eq:HCRB_pol}. The measurement I will mainly consider is a standard Stokes measurement \cite{Kikuchi2020}, presented schematically in Fig.~\ref{fig:stokes}(a). The setup first uniformly splits the signal into three separate light beams. The first beam then travels undisturbed while the second one passes through a $45^\circ$ polarization rotator, turning the polarization to a diagonal-antidiagonal basis. In the case of the third beam the rotator is additionally preceded by a quater-waveplate which sets the polarization basis to a left and right circular. After these operations each beam is separately split by a polarizing beam splitter and photon number resolving detectors measure numbers of photons in each output arm of the respective system.

In order to find what is the best precision attained by a given measurement one can calculate the Fisher information matrix in \eqnref{eq:class_FI} and evaluate the cost.  The bound on estimation precision utilizing the Stokes receiver is given by
\begin{equation}\label{eq:stokes_bound}
\Delta S_1^2+\Delta S_2^2 +\Delta S_3^2\geq \frac{11}{2}S_0-\frac{9}{2S_0}\left(\frac{1}{S_1^2}+\frac{1}{S_2^2}+\frac{1}{S_3^2}\right)^{-1},
\end{equation}
and the detailed calculations can be found in appendix~\ref{app:crb}. It can be seen in Fig.~\ref{fig:kulki}(a) that the expression in \eqnref{eq:stokes_bound} lies between the values of HCRB $\Delta\vec{S}^2=5S_0$ from \eqnref{eq:HCRB_pol} attained for $S_1=S_2=S_3=S_0/\sqrt{3}$ and $\Delta\vec{S}^2=\frac{11}{2}S_0$ which can be obtained when one of the coordinates $S_1,S_2,S_3$ vanishes. Clearly, the bound for Stokes measurement depends on the Stokes parameters and one can achieve the HCRB only locally. This can be in principle overcome if one allows for an additional rotation operation in front of the receiver which would adjust the Stokes vector to have all three coordinates equal. However, from a practical point of view this would require prior knowledge of the signal polarization which means that only small deviations from an otherwise known value can be measured. Note that such parameter dependence of the measurement is usually a feature of general POVMs optimizing the HCRB bound for most estimation problems \cite{Demkowicz2020}. On the other hand even in the worst case scenario pure Stokes receiver attains precision worse than HCRB by only 10\% so the prior knowledge is rather not critical.

In the scenario when the total signal power is fixed the conventional Stokes receiver attains precision lower bounded by
\begin{equation}\label{eq:Stokes_const}
\Delta\vec{S}^2\geq \frac{9}{2S_0}\frac{\left(1+\frac{S_3^2}{S_2^2}\right)\left(1+\frac{S_2^2}{S_1^2}\right)\left(1+\frac{S_1^2}{S_3^2}\right)}{\frac{1}{S_1^2}+\frac{1}{S_2^2}+\frac{1}{S_3^2}}.
\end{equation}
It can be seen in Fig.~\ref{fig:kulki} that for $S_1=S_2=S_3=S_0/\sqrt{3}$ the bound in \eqnref{eq:Stokes_const} saturates HCRB in \eqnref{eq:HCRB_const}. On the other hand whenever one of the Stokes coordinates is vanishing \eqnref{eq:Stokes_const} attains its worst value $9S_0/2$, worse by $12.5\%$ than the corresponding HCRB. Interestingly, note that even the worst case scenario still predicts better precision than the optimal HCRB in the unconstrained case in \eqnref{eq:HCRB}. Knowledge of the total signal power can therefore considerably increase the polarization measurement performance.

Importantly, it was recently shown \cite{Mecozzi2021} that the Stokes receiver, together with a particular estimation procedure can indeed reach the value of $5S_0$ and $4S_0$ precision locally in the general and constrained power cases respectively. Based on this it was conjectured that these values represent quantum limits on polarization measurements. However, the employed estimator in the general scenario attains only $7S_0$ in the worst case instance, which does not saturate the classical Cramer-Rao bound for the Stokes measurement. Although it is known that CRB can be always saturated  by the maximum likelihood estimator \cite{Kay1993} the question remains if there exist any simple estimation procedure that would allow to attain the bounds in \eqnsref{eq:stokes_bound}{eq:Stokes_const} everywhere.

\begin{figure}
	\centering
	\includegraphics[width=1.0\columnwidth]{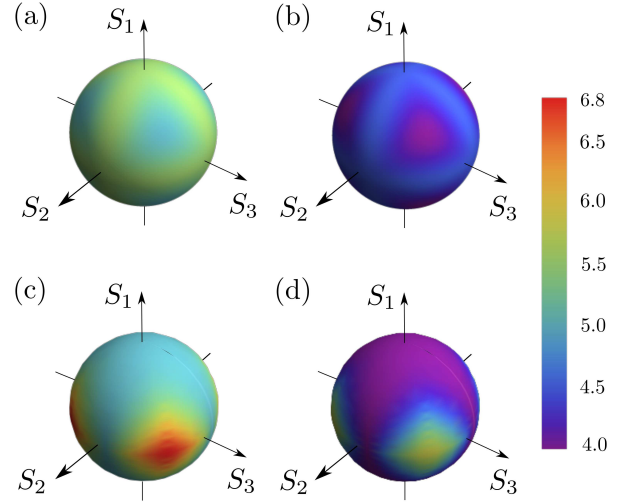}
	\caption{Precision attained as a function of the Stokes vector direction for Stokes receiver in the general case (a), Stokes receiver in the case of known total power $S_0$ (b), tetrahedron receiver in the general case (c), tetrahedron receiver for the known $S_0$ (d). }
\label{fig:kulki}
\end{figure}

Another type of measurement scheme frequently considered in the context of polarization measurement that may seem promising is the so called tetrahedron measurement. This is inspired by a single photon quantum tomography case in which quantum states can be characterized in the Bloch ball picture, isomorphic to the Poincare sphere for polarization \cite{Rehacek2004, Ling2006}. In tetrahedron measurement, presented schematically in Fig.~\ref{fig:stokes}(b), the incoming light is first divided on a partially polarizing beam splitter (PPBS) such that the state evolves according to $\ket{\alpha_H}\ket{\alpha_V}\to(\ket{x\alpha_H}\ket{y\alpha_V})\otimes(\ket{y\alpha_H}\ket{x\alpha_V})$, where $x,y$ are the transmissivity and reflectivity of the beam splitter and brackets indicate separate beams leaving the device. The first beam is then rotated by $45^\circ$ to the diagonal-antidiagonal basis and the second one travels both through a quarter wave plate and a $45^\circ$ rotator in order to end up in right-left circular polarized basis. Each beam is then split by a polarizing beam splitter and one performs photon number measurements on the resulting output ports. Importantly, one can optimize this scheme over the coefficients $x,y$ of the first beam splitter in order to get the optimal performance. The resulting precision bound is quite complicated, but it can be found that
\begin{equation} \label{eq:stokes_tetra}
5S_0\leq\Delta S_1^2+\Delta S_2^2+\Delta S_3^2\leq (4+2\sqrt{2}) S_0
\end{equation}
in the general case and
\begin{equation}\label{eq:HCRB_tet}
4S_0 \leq \Delta S_1^2+\Delta S_2^2+\Delta S_3^2\leq (3+2\sqrt{2})S_0
\end{equation}
for the constrained total power scenario. In both instances the minimum is attained when one of the Stokes coordinates is equal to $S_0$ and others vanish whereas the maximal value is obtained for $S_1=0, S_2=\pm S_0/\sqrt{2}, S_3=\pm S_0/\sqrt{2}$. Note that \eqnsref{eq:stokes_tetra}{eq:HCRB_tet} assume optimal transmission coefficients $x,y$ of the first beam splitter for each value of the Stokes parameters. For both considered polarization estimation scenarios such an optimized tetrahedron detector can locally saturate respective HCRBs in \eqnref{eq:HCRB_pol} and \eqnref{eq:HCRB_const}. Moreover, as can be seen in Fig.~\ref{fig:kulki}(c)(d), it offers precision very close to the ultimate bounds for a considerably larger region of Stokes coefficients than the Stokes receiver, although with a caveat of weaker performance in the worst case scenario. Importantly, one can also saturate respective HCRBs using tetrahedron detection scheme with a fixed non-optimized value of $x$, although this can be accomplished just for a limited range of Stokes parameters.

\section{HCRB in the presence of reference frame}

In most conventional scenarios one usually does not have knowledge about the global phase of the signal. This is because the phase is a relative concept and requires presence of some external phase reference, such as local oscillator, in order to be physically meaningful \cite{Molmer1997}. However, in the context of polarization estimation an interesting question is what would be the bound on precision of Stokes vector measurement had prior knowledge on the global phase been included. An example of such situation would be if one had access to additional reference beam and performed some kind of coherent measurement on a small fraction of the light before the actual measurement of Stokes coefficients. This would enable global phase estimation which could then be used to boost Stokes vector estimation. 

Let me assume that the experimentalist knows the global phase. In such case the state entering the detection apparatus can be described by only three parameters $\{|\alpha_H|,|\alpha_V|,\varphi_-\}$ as $\ket{\psi}=\ket{|\alpha_H|e^{i\varphi_-/2}}\ket{|\alpha_V|e^{-i\varphi_-/2}}$, where I choose global phase to be equal to $\varphi_+=0$. Since global phase is known to the receiver and one estimates all other quantities describing the state there is no nuisance parameter. Therefore in order to find HCRB it is necessary to perform calculations without additional constraint in \eqnref{eq:XH0const1}. The resulting matrices in \eqnref{eq:HCRB} have free variables over which they have to be optimized, resulting in a complicated expression for HCRB. Numerical calculations show that HCRB takes values in the range given by the inequality
\begin{equation}\label{eq:HCRB_phase}
2S_0\leq \Delta S_1^2+\Delta S_2^2+\Delta S_3^2\leq  5S_0.
\end{equation}
The left hand side value is saturated for $S_2=S_3=0$, i.e. horizontally or vertically polarized light, whereas the maximum value on the right hand side is attained for $S_1=0$, i.e. Stokes vector lying in a plane spanned by $S_2,S_3$. If the total power is known in advance the above bound can be improved by an additional factor of $S_0$ subtracted on both sides.

\begin{figure}
	\centering
	\includegraphics[width=\columnwidth]{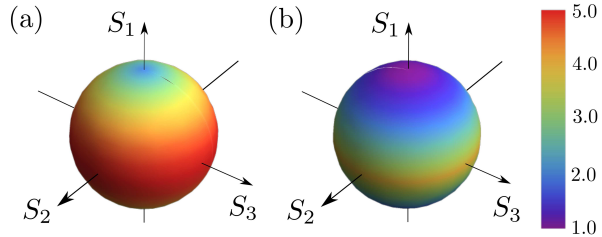}
	\caption{HCRB for Stokes vector estimation as a function of Stokes vector direction in the presence of prior knowledge about the global phase in the general case of unknown $S_0$ (a) and known $S_0$ (b).}
	\label{fig:hcrb_phase}
\end{figure}

It is seen in Fig.~\ref{fig:hcrb_phase} that precision in the presence of prior knowledge about global phase posses a cylindrical symmetry and depends just on the Stokes vector inclination. Interestingly, for some directions of the Stokes vector, the bound is substantially improved with respect to the case of unknown global phase. A similar effect is present in quantum interferometry, where it is known that presence of an external reference beam, serving as a global phase reference, can change achievable bounds on precision of the relative phase delay inside the interferometer \cite{Jarzyna2012, Ataman2020}. Note also, that HCRB in \eqnref{eq:HCRB_phase} in general is not equal to corresponding QCRB in appendix \ref{app:qfi} except at the poles $S_1=\pm S_0$. This means that at these two points one is able to saturate HCRB with conventional measurement in the eigenbasis of SLD. Finally, in contrast to previous cases described in Sec.~\ref{sec:stokes}, classical Stokes receiver or tetrahedron measurement cannot saturate HCRB bound in \eqnref{eq:HCRB_phase}. This is because measurement of the number of photons, which is eventually performed in each arm of these receivers, is insensitive to the global phase. One is therefore forced to use a more sophisticated measurement utilizing a reference beam.

\section{Conclusion}

In conclusion I have showed that the ultimate quantum limit on Stokes vector estimation is equal to $5S_0$, where $S_0$ is the average number of photons in the signal. This bound can be attained locally using conventional Stokes or tetrahedron receivers utilizing photon number resolving detectors. In the case when the total power of the signal is known in advance the bound can be improved to $4S_0$ which is also saturable with Stokes and tetrahedron receivers. Interestingly, these bounds apply in the case in which global phase is unknown to the experimentalist, even though Stokes coefficients do not depend on this parameter. In the opposite regime in which the global phase is an element of prior knowledge the quantum precision bound depends on the signal polarization and in the best case scenario can be improved up to $2S_0$ in the general case and $S_0$ when the total power is known. These latter limits, however, make any physical sense only in the presence of an external reference frame. An interesting question is if there exist any realistic procedure that allows to attain these bounds. Another important problem that remains to be solved is to find the quantum limit on polarization estimation precision in the case of non fully polarized light and for more general quantum states.

\acknowledgements
I acknowledge insightful discussions with R. Demkowicz-Dobrza\'{n}ski and M. Karpi\'{n}ski. This work was supported by the Foundation for Polish Science under the ”Quantum Optical Technologies” project carried out within the International Research Agendas programme co-financed by the European Union under the European Regional Development Fund.

\appendix

\section{HCRB for polarization estimation}
\label{app:hcrb}

In order to find the HCRB for polarization measurement it is convenient to parametrize the quantum states by global and relative phases and absolute values of the amplitudes $\ket{|\alpha_H| e^{\frac{i}{2}(\varphi_++\varphi_-)}}\otimes\ket{|\alpha_V| e^{\frac{i}{2}(\varphi_+-\varphi_-)}}$. As noted in sec.~\ref{sec:hcrb}, the parameters one wants to estimate are $\pmb{\theta}=(|\alpha_H|,|\alpha_V|,\varphi_-)$ with $\varphi_+$ being the nuisance parameter that affects the HCRB bound due to noncommutativity of optimal observables. A first step to obtain the matrix $Z[X]$ in \eqnref{eq:HCRB} is to find the orthonormal basis in which matrices $X_j$ can be written. In general, if the state $\rho=\ket{\psi}\bra{\psi}$ is pure, as is the case for coherent states considered in this paper, the constraints on $X_j$ are given by
\begin{equation}\label{eq:X_constr}
\bra{\psi}X_j\ket{\psi}=0,\quad \bra{\partial_{\theta_j}\psi}X_k\ket{\psi}+\bra{\psi}X_k\ket{\partial_{\theta_j}\psi}=\delta_{jk},
\end{equation}
where $\ket{\partial_{\theta_j}\psi}=\partial_{\theta_j}\ket{\psi}$ denotes the derivative of state with respect to parameter $\theta_j$. Eq.~(\ref{eq:X_constr}) suggests that a useful orthonormal basis is the one characterizing the subspace spanned by vectors $\ket{\psi}$ and $\{\ket{\partial_{\theta_j}\psi}\}$. In order to find this basis one needs to perform Gram-Schmidt orthogonalization of the set of vectors spanning this subspace. Taking $\ket{\psi}=\ket{\alpha_H}\otimes\ket{\alpha_V}$ and parameters $|\alpha_H|,|\alpha_V|,\varphi_-,\varphi_+$ one obtains then the basis
\begin{gather}\label{eq:e0}
\ket{e_0}=\ket{\alpha_H}\otimes\ket{\alpha_V},\\
\ket{e_1}=\partial_{|\alpha_H|}(\ket{\alpha_H}\otimes\ket{\alpha_V})=\ket{\partial_{|\alpha_H|}\alpha_H}\otimes\ket{\alpha_V},\\
\ket{e_2}=\partial_{|\alpha_V|}(\ket{\alpha_H}\otimes\ket{\alpha_V})=\ket{\alpha_H}\otimes\ket{\partial_{|\alpha_V|}\alpha_V}.\label{eq:e2}
\end{gather}
Note that this basis has only three elements, meaning that the subspace spanned by $\{\ket{\alpha_H}\otimes\ket{\alpha_V},\partial_{|\alpha_H|}(\ket{\alpha_H}\otimes\ket{\alpha_V}),\partial_{|\alpha_V|}(\ket{\alpha_H}\otimes\ket{\alpha_V}),\partial_{\varphi_-}(\ket{\alpha_H}\otimes\ket{\alpha_V}),\partial_{\varphi_+}(\ket{\alpha_H}\otimes\ket{\alpha_V})\}$ is actually three dimensional. Therefore, vectors $\partial_{\varphi_-}(\ket{\alpha_H}\otimes\ket{\alpha_V})$ and $\partial_{\varphi_+}(\ket{\alpha_H}\otimes\ket{\alpha_V})$ are linear combinations of vectors found in \eqnsref{eq:e0}{eq:e2}
\begin{multline}
\partial_{\varphi_\pm}(\ket{\alpha_H}\otimes\ket{\alpha_V})=\\=\frac{i}{2}(|\alpha_H|^2\pm|\alpha_V|^2)\ket{e_0}+\frac{i|\alpha_H|}{2}\ket{e_1}\pm\frac{i|\alpha_V|}{2}\ket{e_2}.
\end{multline}
The constraints on $X_{|\alpha_H|}$ due to derivatives with respect to the estimated parameters $|\alpha_H|,\,|\alpha_V|,\varphi_-$ from \eqnref{eq:X_constr} can be therefore written as
\begin{align}
&(X_{|\alpha_H|})_{00}=0,\quad 2\textrm{Re}(X_{|\alpha_H|})_{01}=1, \quad 2\textrm{Re}(X_{|\alpha_H|})_{02}=0, \nonumber\\ &|\alpha_H|\textrm{Im}(X_{|\alpha_H|})_{01}-|\alpha_V|\textrm{Im}(X_{|\alpha_H|})_{02}=0.
\label{eq:cons}
\end{align}
Based on the above equations operator $X_{|\alpha_H|}$ has a following form
\begin{equation}\label{eq:X_H0}
X_{|\alpha_H|}=\left(
\begin{array}{c c c}
0 & \frac{1}{2}+ib & i\frac{|\alpha_H|}{|\alpha_V|}b\\
\frac{1}{2}-ib & 0 & 0\\
-i\frac{|\alpha_H|}{|\alpha_V|}b & 0 &0
\end{array}\right),
\end{equation}
where $b\in\mathbb{R}$ is a free parameter over which one has to optimize the HCRB. The matrix elements not constrained by \eqnref{eq:cons} were set to $0$ since they do not enter the expression for $Z[X]$. As mentioned above, the global phase, as a nuisance parameter, also has an impact on the form of $X_j$ due to noncommutativity. This manifests as an additional constraint corresponding to a derivative with respect to $\varphi_+$ which reads
\begin{equation}\label{eq:XH0const_ap}
|\alpha_H|\textrm{Im}(X_{|\alpha_H|})_{01}+|\alpha_V|\textrm{Im}(X_{|\alpha_H|})_{02}=0.
\end{equation}
The above equation further constraints the form of $X_{|\alpha_H|}$ in \eqnref{eq:X_H0} and a unique solution can be found, equal to
\begin{equation}\label{eq:X_H}
X_{|\alpha_H|}=\left(
\begin{array}{c c c}
0 & \frac{1}{2} & 0\\
\frac{1}{2} & 0 & 0\\
0 & 0 &0
\end{array}\right),
\end{equation}
with $b=0$ in \eqnref{eq:X_H0}. Using similar method, one can obtain that operators $X_{|\alpha_V|}$ and $X_{\varphi_-}$ are given by
\begin{equation}\label{eq:X_V}
X_{|\alpha_V|}=\left(
\begin{array}{c c c}
0 & 0 & \frac{1}{2}\\
0 & 0 & 0\\
\frac{1}{2} & 0 &0
\end{array}\right),
\end{equation}
and
\begin{equation}\label{eq:X_ph}
X_{\varphi_-}=\left(
\begin{array}{c c c}
0 & -\frac{i}{2|\alpha_H|} & \frac{i}{2|\alpha_V|}\\
\frac{i}{2|\alpha_H|} & 0 & 0\\
-\frac{i}{2|\alpha_V|} & 0 &0
\end{array}\right).
\end{equation}
Note that none of the operators $X_j$ posses any kind of free parameter over which one could perform optimization in HCRB. This is because the constraint originating from the lack of knowledge about global phase $\Tr[\partial_{\varphi_+}\rho X_j]=0$ fixes any kind of freedom in the choice of $X_j$'s.

Using \eqnsref{eq:X_H}{eq:X_ph} the matrix $Z[X]$ is equal to
\begin{equation}\label{eq:zx}
Z[X]=\frac{1}{4}\left(
\begin{array}{c c c}
1 & 0 & \frac{i}{|\alpha_H|}\\
0 & 1 & -\frac{i}{|\alpha_V|}\\
-\frac{i}{|\alpha_H|} & \frac{i}{|\alpha_V|} &\frac{1}{|\alpha_H|^2}+\frac{1}{|\alpha_V|^2}
\end{array}\right).
\end{equation}
In order to find the HCRB bound for polarization one can now transform \eqnref{eq:zx} through Jacobian of the transformation in eqs.~(\ref{eq:stokes_parameter_ph3}) of the main text
\begin{equation}\label{eq:jacobian}
J=2\left(
\begin{array}{c c c}
|\alpha_V|\cos\varphi_- & |\alpha_H|\cos\varphi_-  & -|\alpha_H||\alpha_V|\sin\varphi_- \\
|\alpha_V|\sin\varphi_-  & |\alpha_H|\sin\varphi_-  & |\alpha_H||\alpha_V|\cos\varphi_-\\
|\alpha_H| & -|\alpha_V| & 0
\end{array}\right),
\end{equation}
resulting in 
\begin{equation}\label{eq:zxj}
JZ[X]J^T=\left(
\begin{array}{c c c}
S_0 & -i S_1 & iS_3\\
i S_1 & S_0 & -iS_2\\
- iS_3& iS_2 &S_0
\end{array}\right).
\end{equation}
Plugging \eqnref{eq:zxj} into HCRB one obtains that precision is lower bounded by
\begin{equation}\label{eq:HCRB_pol_ap}
\Delta S_1^2+\Delta S_2^2 +\Delta S_3^2\geq 5S_0.
\end{equation}
In order to find the HCRB in the presence of prior knowledge about the global phase the only change in the above reasoning is to drop the constraint in \eqnref{eq:XH0const_ap}. In such a case the resulting matrices $X_j$ will have free variables, such as in \eqnref{eq:X_H0}, over which one has to optimize in \eqnref{eq:HCRB}.

\section{HCRB for constrained total power}
\label{app:hcrb_s0}

If the total power of light, $S_0$, is known a priori, one can use the relation $|\alpha_V|=\sqrt{S_0-|\alpha_H|^2}$ which means that there are only two parameters to estimate $(|\alpha_H|,\,\varphi_-)$. Similarly as in the general case, in order to find HCRB in the first step one needs to find an orthonormal basis of the space spanned by $\ket{\psi},\,\ket{\partial_{|\alpha_H|}\psi},\,\ket{\partial_{\varphi_-}\psi},\,\ket{\partial_{\varphi_+}\psi}$ which is equal to
\begin{gather}\label{eq:e0_const}
\ket{e_0}=\ket{\psi},\\
\ket{e_1}=\sqrt{\beta}\ket{\partial_{|\alpha_H|}\psi},\\
\ket{e_2}=\frac{\sqrt{S_0}}{\gamma}\left[\ket{\partial_{\varphi_-}\psi}-i\gamma\ket{\psi}-i\beta|\alpha_H|\ket{\partial_{|\alpha_H|}\psi}\right],\label{eq:e2_const}
\end{gather}
where $\beta=1-\frac{|\alpha_H|^2}{S_0}$ and $\gamma=|\alpha_H|^2-\frac{S_0}{2}$. As can be seen this subspace is three dimensional and the derivative with respect to the global phase can be expressed by linear combination of vectors in \eqnsref{eq:e0_const}{eq:e2_const}
\begin{equation}
\ket{\partial_{\varphi_+}\psi}=\frac{1}{2}\left(\sqrt{S_0}\ket{e_2}+iS_0\ket{e_0}\right).
\end{equation}
Therefore, the constraints on $X_{|\alpha_H|}$ originating from the estimated parameters are given by
\begin{align}
(X_{|\alpha_H|})_{00}=0,\quad 2\textrm{Re}(X_{|\alpha_H|})_{01}=\sqrt{\beta}, \nonumber\\
\frac{\gamma}{\sqrt{S_0}}\textrm{Re}(X_{|\alpha_H|})_{02}-\sqrt{\beta}|\alpha_H|\textrm{Im}(X_{|\alpha_H|})_{01}=0,
\end{align}
whereas the constraint coming from the global phase reads
\begin{equation}\label{eq:const_s0}
2\textrm{Re}(X_{|\alpha_H|})_{02}=0.
\end{equation}
A general solution to these equations is given by
\begin{equation}\label{eq:X_H_S0}
X_{|\alpha_H|}=\left(
\begin{array}{c c c}
0 & \frac{\sqrt{\beta}}{2} & -i b\\
\frac{\sqrt{\beta}}{2} & 0 & 0\\
i b & 0 &0
\end{array}\right),
\end{equation}
where $b\in\mathbb{R}$ is a free parameter. Similarly one obtains
\begin{equation}\label{eq:X_p_S0}
X_{\varphi_-}=\left(
\begin{array}{c c c}
0 & -\frac{i}{2|\alpha_H|\sqrt{\beta}} & -i c\\
\frac{i}{2|\alpha_H|\sqrt{\beta}} & 0 & 0\\
i c & 0 &0
\end{array}\right),
\end{equation}
where once again $c\in\mathbb{R}$ is a free parameter. The resulting $Z[X]$ matrix in HCRB is equal to
\begin{equation}
Z[X]=\left(
\begin{array}{c c}
\frac{\beta}{4}+b^2 & \frac{i}{4|\alpha_H|} +bc\\
-\frac{i}{4|\alpha_H|}+bc & \frac{1}{4|\alpha_H|^2\beta}+c^2
\end{array}\right).
\end{equation}
The transformation between Stokes coordinates and coherent states parameters is given by
\begin{equation}
|\alpha_H|=\sqrt{\frac{S_0+S_1}{2}},\quad
\varphi_-=\arctan\frac{S_3}{S_2},
\end{equation}
which results in a Jacobian
\begin{equation}\label{eq:jacobian_s0}
J=2\left(
\begin{array}{c c}
\frac{S_0-2|\alpha_H|^2}{\sqrt{S_0-|\alpha_H|^2}}\cos\varphi_- & -|\alpha_H|\sqrt{S_0-|\alpha_H|^2}\sin\varphi_-\\
\frac{S_0-2|\alpha_H|^2}{\sqrt{S_0-|\alpha_H|^2}}\sin\varphi_- & |\alpha_H|\sqrt{S_0-|\alpha_H|^2}\cos\varphi_-\\
2|\alpha_H| & 0
\end{array}\right).
\end{equation}
After reparametrization the HCRB has to be optimized over $b$ and $c$ resulting in optimal values $b=c=0$. The precision bound is then given by
\begin{equation}\label{eq:HCRB_const_app}
\Delta S_1^2+\Delta S_2^2+\Delta S_3^2\geq4S_0.
\end{equation}
Similarly as in app.~\ref{app:hcrb}, in order to find the HCRB in the presence of prior knowledge about the global phase, one needs to repeat the calculations without the constraint in \eqnref{eq:const_s0} and optimize the resulting expression in \eqnref{eq:HCRB} over all free variables.

\section{Quantum Cramer-Rao bounds}

\label{app:qfi}

In order to find QCRB one needs to first calculate SLDs for all parameters. In a general quantum multiparameter estimation model with pure states $\ket{\psi_{\pmb{\theta}}}$ SLD for the $j$-th parameter is equal to
\begin{equation}
L_{\theta_j}=2(\ket{\partial_{\theta_j}\psi}\bra{\psi}+\ket{\psi}\bra{\partial_{\theta_j}\psi}).
\end{equation}
Assuming one posses knowledge or access to the global phase, the state in the polarization measurement problem is given by $\ket{\psi_{\pmb{\theta}}}=\ket{|\alpha_H|e^{i\varphi_-/2}}\ket{|\alpha_V|e^{-i\varphi_-/2}}$, where I have chosen global phase to be equal to $\varphi_+=0$. One can easily calculate resulting SLDs which leads to a following quantum Fisher information matrix
\begin{equation}
\mathcal{Q}=\left(
\begin{array}{c c c}
4 &0 & 0\\
0 & 4 & 0\\
0 & 0 & S_0
\end{array}\right)
\end{equation}
in the $(|\alpha_H|,|\alpha_V|,\varphi_-)$ parametrization. The resulting cost, after transforming to Stokes coefficients parametrization reads
\begin{equation}\label{eq:QCRB_phase}
\Tr(\mathcal{Q}^{-1})=3S_0-\frac{S_1^2}{S_0},
\end{equation}
which takes the values in the region $2S_0\leq \Tr(\mathcal{Q}^{-1})\leq 3S_0 $. The minimal value is obtained for $S_1=S_0$ whereas the worst one for $S_1=0$. The QCRB bound, however, cannot be saturated in this case as $\Tr(\rho_{\pmb{\theta}}[L_{j},L_k])\neq 0$ for $j=\varphi_-$ and $k=|\alpha_H|,\,|\alpha_V|$. In particular
\begin{gather}
\Tr(\ket{\psi_{\pmb{\theta}}}\bra{\psi_{\pmb{\theta}}}[L_{\varphi_-},L_{|\alpha_H|}])=-2i|\alpha_H|,\\
\Tr(\ket{\psi_{\pmb{\theta}}}\bra{\psi_{\pmb{\theta}}}[L_{\varphi_-},L_{|\alpha_V|}])=2i|\alpha_V|.
\end{gather}
Note that this in principle means that HCRB is different than QCRB for all possible values of amplitudes, except $|\alpha_H|=|\alpha_V|=0$, which is a vacuum state. On the other hand, it is seen in Fig.~\ref{fig:kulki} that on the poles of the Poincare sphere $S_1=\pm S_0$ the respecitve bounds coincide. The reason for this apparent contradiction is that exactly at these two points one has either $|\alpha_V|=0$ (northern pole) or $|\alpha_H|=0$ (sourthern pole) and the Stokes coefficients in \eqnref{eq:stokes_parameter_ph3} do not depend on the realitve phase in any way. Therefore, the respective contribution for the relative phase estimation does not enter the expressions for both bounds and one can perform estimation neglecting this parameter, meaning that only SLDs for absolute values of both amplitudes matter. These two operator commute, therefore, QCRB is equal to HCRB on poles of the Poincare sphere.

In a realistic scenario, when one does not have access to an external reference frame, the state should be averaged over the global phase \cite{Molmer1997, Jarzyna2012}. In such a case one obtains
\begin{equation}
\rho_{\pmb{\theta}}=\int\frac{d\varphi_+}{2\pi}\ket{\psi_{\pmb{\theta}}}\bra{\psi_{\pmb{\theta}}}
=\sum_{N=0}^\infty p_N(\pmb{\theta}) \ket{v^N_{\pmb{\theta}}}\bra{v^N_{\pmb{\theta}}},
\end{equation}
where
\begin{gather}
p_N(\pmb{\theta})=e^{-S_0}\frac{S_0^N}{N!},\\
\ket{v^N_{\pmb{\theta}}}=\frac{1}{S_0^{N/2}}\sum_{n=0}^N\sqrt{{N}\choose{n}}|\alpha_H|^n|\alpha_V|^{N-n}e^{i\varphi_-n}\ket{n,N-n}.
\end{gather}
Note, that different $\ket{v^N_{\pmb{\theta}}}$ are orthogonal $\braket{v^N_{\pmb{\theta}}}{v^M_{\pmb{\theta}}}=\delta_{N,M}$ as they live in subspaces with different total numbers of photons. One can evaluate SLD's directly and obtain
\begin{multline}
L_{|\alpha_H|}=2|\alpha_H|\sum_{N=0}^\infty\left(\frac{N}{S_0}-1\right)\ket{v^N_{\pmb{\theta}}}\bra{v^N_{\pmb{\theta}}}+\\+2\sum_{N=0}^\infty\left(\ket{\partial_{|\alpha_H|}v^N_{\pmb{\theta}}}\bra{v^N_{\pmb{\theta}}}+\ket{v^N_{\pmb{\theta}}}\bra{\partial_{|\alpha_H|}v^N_{\pmb{\theta}}}\right),
\end{multline}
\begin{multline}
L_{|\alpha_V|}=2|\alpha_V|\sum_{N=0}^\infty\left(\frac{N}{S_0}-1\right)\ket{v^N_{\pmb{\theta}}}\bra{v^N_{\pmb{\theta}}}+\\+2\sum_{N=0}^\infty\left(\ket{\partial_{|\alpha_V|}v^N_{\pmb{\theta}}}\bra{v^N_{\pmb{\theta}}}+\ket{v^N_{\pmb{\theta}}}\bra{\partial_{|\alpha_V|}v^N_{\pmb{\theta}}}\right)
\end{multline}
and
\begin{equation}
L_{\varphi_-}=2\sum_{N=0}^\infty\left(\ket{\partial_{\varphi_-}v^N_{\pmb{\theta}}}\bra{v^N_{\pmb{\theta}}}+\ket{v^N_{\pmb{\theta}}}\bra{\partial_{\varphi_-}v^N_{\pmb{\theta}}}\right).
\end{equation}
The resulting quantum Fisher information matrix is equal to
\begin{equation}
\mathcal{Q}=\left(
\begin{array}{c c c}
4 &0 & 0\\
0 & 4 & 0\\
0 & 0 & \frac{4|\alpha_H|^2|\alpha_V|^2}{S_0}
\end{array}\right),
\end{equation}
which, after transformation into Stokes coefficients parametrization gives the cost equal to
\begin{equation}
\Tr(\mathcal{Q}^{-1})=3S_0.
\end{equation}
 
Similarly to the previous case, in the phase averaged scenario SLDs do not commute on the state. One obtains
\begin{gather}
\Tr(\ket{\psi_{\pmb{\theta}}}\bra{\psi_{\pmb{\theta}}}[L_{\varphi_-},L_{|\alpha_H|}])=-\frac{8i|\alpha_H||\alpha_V|^2}{S_0},\\
\Tr(\ket{\psi_{\pmb{\theta}}}\bra{\psi_{\pmb{\theta}}}[L_{\varphi_-},L_{|\alpha_V|}])=-\frac{8i|\alpha_V||\alpha_H|^2}{S_0}.
\end{gather}
Therefore, QCRB for the phase averaged state is also unattainable.

Finally, one can also consider scenario with known total power $S_0$, where the estimated parameters are $(|\alpha_H|,\varphi_-)$. The resulting expressions for quantum Fisher information matrix when one has access to the global phase is given by
\begin{equation}
\mathcal{Q}=\left(
\begin{array}{c c}
\frac{8S_0}{S_0-S_1} &0 \\
0 & S_0 
\end{array}\right),
\end{equation}
which, after transformation to Stokes coefficients parametrization gives precision equal to
\begin{equation}
\Delta S_1^2+\Delta S_2^2+\Delta S_3^2=2S_0-\frac{S_1^2}{S_0}.
\end{equation}
which takes values in the interval $[S_0,2S_0]$, the minimal obtained for $S_1=S_0$ and maximal for $S_1=0$. For the phase averaged scenario the respective quantities are given by
\begin{equation}
\mathcal{Q}=\left(
\begin{array}{c c}
\frac{8S_0}{S_0-S_1} &0 \\
0 & S_0-\frac{S_1^2}{S_0} 
\end{array}\right),
\end{equation}
which, after transformation to Stokes coefficients parametrization gives precision equal to
\begin{equation}
\Delta S_1^2+\Delta S_2^2+\Delta S_3^2=2S_0.
\end{equation}
In both cases SLDs do not commute on the respective states.

\section{Classical Fisher information matrix for Stokes receiver}

\label{app:crb}

The quantum states of light impinging on photon number detectors in each arm of the Stokes receiver  in Fig.~\ref{fig:stokes}(a) are coherent states with estimated parameters encoded in their amplitude. The photon number distribution of a coherent state with amplitude $|\alpha_{\pmb{\theta}}|=\sqrt{\bar{n}_{\pmb{\theta}}}$ is a Poisson distribution
\begin{equation}
p_k(\pmb{\theta})=e^{-\bar{n}_{\pmb{\theta}}}\frac{\bar{n}_{\pmb{\theta}}^k}{k!}.
\end{equation}
Plugging this expression into the formula for Fisher information matrix in \eqnref{eq:class_FI} gives a following expression for its elements
\begin{equation}
F_{jl}=\sum_{k=0}^\infty \frac{1}{p_k(\pmb{\theta})}\left(\frac{\partial p_k(\pmb{\theta})}{\partial \bar{n}_{\pmb{\theta}}}\right)^2\frac{\partial\bar{n}_{\pmb{\theta}}}{\partial \theta_j}\frac{\partial\bar{n}_{\pmb{\theta}}}{\partial \theta_l}=\frac{1}{\bar{n}_{\pmb{\theta}}}\frac{\partial\bar{n}_{\pmb{\theta}}}{\partial \theta_j}\frac{\partial\bar{n}_{\pmb{\theta}}}{\partial \theta_l}.
\end{equation}
The amplitudes of light impinging on each detector are given by
\begin{equation}\label{eq:amplitudes}
\tilde{\alpha}_{H,V}=\frac{\alpha_{H,V}}{\sqrt{3}},\quad\tilde{\alpha}_\pm=\frac{\alpha_H\pm\alpha_V}{\sqrt{6}},\quad \tilde{\alpha}_{R,L}=\frac{\alpha_H\pm i\alpha_V}{\sqrt{6}},\\
\end{equation}
in the first, second and third arm respectively, where the factor $1/\sqrt{3}$ comes from the division of the signal on a three-way splitter.

The Fisher information matrices for each of the three arms of the Stokes receiver are given as sums of Fisher matrices on both detectors in these respective arms. For estimation of $|\alpha_H|,|\alpha_V|,\varphi_-$ they read
\begin{equation}\label{eq:FI_HV}
\mathcal{F}^{HV}=\left(
\begin{array}{c c c}
\frac{4}{3} &0 & 0\\
0 & \frac{4}{3} & 0\\
0 & 0 & 0
\end{array}\right),
\end{equation}
\begin{widetext}
\begin{equation}
\mathcal{F}^{+-}=\frac{1}{3}\left(
\begin{array}{c c c}
\frac{|\alpha_H|^2S_0+|\alpha_V|^2(|\alpha_V|^2-3|\alpha_H|^2)\cos^2\varphi_-}{|\alpha_+|^2|\alpha_-|^2} & |\alpha_H||\alpha_V|\frac{S_0\sin^2\varphi_-}{|`\alpha_+|^2|\alpha_-|^2} & \frac{|\alpha_H||\alpha_V|^2(|\alpha_H|^2-|\alpha_V|^2)}{2|\alpha_+|^2|\alpha_-|^2}\sin2\varphi_-\\
|\alpha_H||\alpha_V|\frac{S_0\sin^2\varphi_-}{|\alpha_+|^2|\alpha_-|^2} & \frac{|\alpha_V|^2S_0+|\alpha_H|^2(|\alpha_H|^2-3|\alpha_V|^2)\cos^2\varphi_-}{|\alpha_+|^2|\alpha_-|^2} & -\frac{|\alpha_V||\alpha_H|^2(|\alpha_H|^2-|\alpha_V|^2)}{2|\alpha_+|^2|\alpha_-|^2}\sin2\varphi_-\\
\frac{|\alpha_H||\alpha_V|^2(|\alpha_H|^2-|\alpha_V|^2)}{2|\alpha_+|^2|\alpha_-|^2}\sin2\varphi_- & -\frac{|\alpha_V||\alpha_H|^2(|\alpha_H|^2-|\alpha_V|^2)}{2|\alpha_+|^2|\alpha_-|^2}\sin2\varphi_- & \frac{|\alpha_H|^2|\alpha_V|^2S_0}{|\alpha_+|^2|\alpha_-|^2}\sin^2\varphi_-
\end{array}\right),
\end{equation}
\end{widetext}
and
\begin{widetext}
\begin{equation}
\mathcal{F}^{RL}=\frac{4}{3}\left(
\begin{array}{c c c}
\frac{|\alpha_H|^2S_0+|\alpha_V|^2(|\alpha_V|^2-3|\alpha_H|^2)\sin^2\varphi_-}{|\alpha_R|^2|\alpha_L|^2} & |\alpha_H||\alpha_V|\frac{S_0\cos^2\varphi_-}{|\alpha_R|^2|\alpha_L|^2} & -\frac{|\alpha_H||\alpha_V|^2(|\alpha_H|^2-|\alpha_V|^2)}{2|\alpha_R|^2|\alpha_L|^2}\sin2\varphi_-\\
|\alpha_H||\alpha_V|\frac{S_0\cos^2\varphi_-}{|\alpha_R|^2|\alpha_L|^2} & \frac{|\alpha_V|^2S_0+|\alpha_H|^2(|\alpha_H|^2-3|\alpha_V|^2)\sin^2\varphi_-}{|\alpha_R|^2|\alpha_L|^2} & \frac{|\alpha_V||\alpha_H|^2(|\alpha_H|^2-|\alpha_V|^2)}{2|\alpha_R|^2|\alpha_L|^2}\sin2\varphi_-\\
-\frac{|\alpha_H||\alpha_V|^2(|\alpha_H|^2-|\alpha_V|^2)}{2|\alpha_R|^2|\alpha_L|^2}\sin2\varphi_- & \frac{|\alpha_V||\alpha_H|^2(|\alpha_H|^2-|\alpha_V|^2)}{2|\alpha_R|^2|\alpha_L|^2}\sin2\varphi_- & \frac{|\alpha_H|^2|\alpha_V|^2S_0}{|\alpha_R|^2|\alpha_L|^2}\cos^2\varphi_-
\end{array}\right).
\label{eq:FI_RL}
\end{equation}
\end{widetext}
Upon adding \eqnsref{eq:FI_HV}{eq:FI_RL} one obtains classical Fisher information for Stokes receiver, which then has to be reperamterized through Jacobian given in \eqnref{eq:jacobian} in order to obtain precision bound for Stokes vector estimation, giving
\begin{equation}
\Delta S_1^2+\Delta S_2^2 +\Delta S_3^2\geq \frac{11}{2}S_0-\frac{9}{2S_0}\left(\frac{1}{S_1^2}+\frac{1}{S_2^2}+\frac{1}{S_3^2}\right)^{-1}.
\end{equation}

Similarly, one can obtain FI matrix for the Stokes receiver in the case of known total signal power which has been analyzed in Sec.~\ref{sec:stokes}(a). Once again, using \eqnref{eq:amplitudes} and \eqnref{eq:class_FI} together with the constraint $|\alpha_V|=\sqrt{S_0-|\alpha_H|^2}$ one gets
\begin{equation}\label{eq:FI_HV_pow}
\mathcal{F}_{HV}=\left(
\begin{array}{c c}
\frac{4}{3(S_0-|\alpha_H|^2)} &0 \\
0 & 0
\end{array}\right),
\end{equation}
\begin{widetext}
\begin{equation}
\mathcal{F}_{+-}=\frac{4}{3}\left(
\begin{array}{c c}
\frac{S_0(S_0-2|\alpha_H|^2)^2\cos^2\varphi_-}{(S_0-|\alpha_H|^2)(S_0^2-4|\alpha_H|^2(S_0-|\alpha_H|^2)\cos^2\varphi_-)} & -\frac{|\alpha_H|S_0(S_0-2|\alpha_H|^2)\sin \varphi_-\cos\varphi_-}{S_0^2-4|\alpha_H|^2(S_0-|\alpha_H|^2)\cos^2\varphi_-}\\
-\frac{|\alpha_H|S_0(S_0-2|\alpha_H|^2)\sin \varphi_-\cos\varphi_-}{S_0^2-4|\alpha_H|^2(S_0-|\alpha_H|^2)\cos^2\varphi_-} & \frac{|\alpha_H|^2S_0(S_0-2|\alpha_H|^2)\sin^2\varphi_-}{S_0^2-4|\alpha_H|^2(S_0-|\alpha_H|^2)\cos^2\varphi_-}
\end{array}\right),
\label{eq:FI_RL_pow}
\end{equation}
\end{widetext}
and
\begin{widetext}
\begin{equation}
\mathcal{F}_{RL}=\frac{4}{3}\left(
\begin{array}{c c}
\frac{S_0(S_0-2|\alpha_H|^2)^2\sin^2\varphi_-}{(S_0-|\alpha_H|^2)(S_0^2-4|\alpha_H|^2(S_0-|\alpha_H|^2)\sin^2\varphi_-)} & \frac{|\alpha_H|S_0(S_0-2|\alpha_H|^2)\sin \varphi_-\cos\varphi_-}{S_0^2-4|\alpha_H|^2(S_0-|\alpha_H|^2)\sin^2\varphi_-}\\
\frac{|\alpha_H|S_0(S_0-2|\alpha_H|^2)\sin \varphi_-\cos\varphi_-}{S_0^2-4|\alpha_H|^2(S_0-|\alpha_H|^2)\sin^2\varphi_-} & \frac{|\alpha_H|^2S_0(S_0-2|\alpha_H|^2)\cos^2\varphi_-}{S_0^2-4|\alpha_H|^2(S_0-|\alpha_H|^2)\sin^2\varphi_-}
\end{array}\right)
\end{equation}
\end{widetext}
for parameters $|\alpha_H|,\varphi_-$ respectively. This has to be then reparametrized through Jacobian in \eqnref{eq:jacobian_s0}, resulting in the bound
\begin{equation}
\Delta\vec{S}^2\geq \frac{9}{2S_0}\frac{\left(1+\frac{S_3^2}{S_2^2}\right)\left(1+\frac{S_2^2}{S_1^2}\right)\left(1+\frac{S_1^2}{S_3^2}\right)}{\frac{1}{S_1^2}+\frac{1}{S_2^2}+\frac{1}{S_3^2}}.
\end{equation}

A similar analysis can be performed for tetrahedron measurement with additional optimization of the Cramer-Rao bound over the transmission and reflection coefficients of the first beam splitter in Fig.~\ref{fig:stokes}(b).

\bibliography{polarizationbiblio}

\end{document}